# Impact of annealing temperature on structural, electrical and optical properties of epitaxial GaN thin films grown on sapphire substrates by PA-MBE

Mahitosh Biswas

Centre for Research in Nanotechnology and Science, IIT Bombay, Powai, Mumbai, India.

We report epitaxial growth and characterization of GaN thin films on sapphire (0001) substrates using low temperature GaN intermediate layer by plasma assisted molecular beam epitaxy (PA-MBE) technique. As grown and annealed GaN thin films were studied by high-resolution X-ray diffraction (HRXRD), atomic force microscopy (AFM), Hall Effect and photoluminescence (PL). It has been found that there is a significant improvement in the quality of the GaN films after annealing at 725°C in terms of electron mobility, full width at half maximum (FWHM) of omega ($\omega$) scan around (0002) XRD peak of GaN films. Screw dislocation density obtained from the FWHM of GaN (0002) $\omega$ scan and etch pitch density calculated from AFM image are $6.4 \times 10^8$ cm$^{-2}$ and $5.1 \times 10^8$ cm$^{-2}$ respectively. In PL measurement, FWHM of near band edge (NBE) peak of GaN films has been found to be 30 meV.

1. Introduction:

III-Nitrides semiconductors like GaN, AlN, InN and their alloys have attracted a great attention in optoelectronics and high temperature andr high power electronic devices owing to their excellent properties such as a wide range of direct band gaps and high temperature stability. Molecular beam epitaxy (MBE) has been one of promising techniques to grow III-nitride based device heterostructures such as light emitting diodes [1,2,3], lased diodes [4,5], solar cell [6,7], UV Photodetectors [8], sensors [9,10] etc. Recently Plasma Assisted Molecular Beam Epitaxy (PAMBE) is extensively used to grow GaN based materials under slightly Ga-rich conditions in order to obtain good quality of the surface morphology, electrical and optical properties. Mostly nitride films are grown on lattice and thermal coefficient mismatched substrates such as Si, SiC and sapphire resulting in stress generation, roughness and defects formation which restrict high electron mobility, high output power and long life time operation in device application. To achieve better control over the crystalline quality with reduced defects and dislocations, homoepitaxial GaN thin film growth offers an advantage to fabricate efficient devices. These defects and dislocations originate from the strain or stress produced in the films. The main source of strain generated in homoepitaxially grown GaN films is the presence of point defects. It is reported in the literature [11,12] that the presence of stress or strain in the films can severely affect the structural and optical properties of GaN films. Moreover, GaN films grown on standard GaN template and two-step epitaxial layer overgrowth (ELO) GaN template exhibit a significant difference in density of surface pits, usually observed in MBE-grown GaN films and associated with the surface termination of threading dislocations [13,14,15] originating from nucleation layer.

Precise control of growth in Ga rich conditions below the droplet region [16] has become the subject of intense research to achieve good quality of GaN films. Koblmüller et al. [17] reported high temperature growth of high quality GaN films indicating the smoothest morphologies near the boundary between N-rich and Ga-rich growth (0.75 < Ga/N < 1.1) on metal-organic chemical vapor deposition (MOCVD) GaN templates by PAMBE. Under Ga rich condition, a metallic Ga wetting adlayer exists on the surface with many advantageous effects, such as enhanced surface adatom diffusion [18], a two-dimensional 2D growth mode and consequently smoother surface morphologies [19,20]. Dislocation density more than $1\times10^8$ cm$^2$/V.sec plays an important role to scattering mechanism in the GaN films [21]. To obtain high mobility of the GaN films, it is suggested to reduce dislocation generated in the films as much as possible. Mobility of the GaN films grown by MBE is typically smaller than that of MOCVD grown films. Powel et al. [22] and Grandjean and Massies [23] reported much lower value of mobility in the range of 100-300 cm$^2$/V.sec.

We report the effect of annealing on structural, electrical and optical properties of GaN thin films grown by PAMBE. In this study, we report structural quality improvement of the annealed GaN films in terms of FWHM of GaN (0002) XRD peak, screw dislocation density calculated from XRD peak and RMS roughness obtained from AFM. Electron mobility has been investigated by Hall Effect and optical properties of the annealed GaN films have been studied by PL measurement in terms of FWHM of the dominant Near Band Edge (NBE) transition. We have seen that ex-situ annealing temperature 725°C is the best suitable temperature for improving the film quality considering the structural, surface morphological, optical and electrical properties.

2. Experimental Methods:

GaN thin films were grown on Sapphire (0001) substrates using low temperature (LT) GaN intermediate layer by PA-MBE system equipped with standard effusion cells for Ga, Al, In and a radio frequency plasma source (ADDON) for supplying active nitrogen. 2 inch sapphire substrates were cleaned by a 7:3 solution of $H_2SO_4$: $H_2O_2$. Substrates were dipped into the solution at 160°C for 30 mins. Then this was followed by a DI water rinse for 3 mins. On the back side of wafers 1 μm Ti has been deposited by electron beam evaporator to provide sufficient heat transfer. These wafers were degassed in preparation chamber at 450°C overnight in order to remove residual contaminants present on the surface of the wafers. Nitridation of the sapphire substrates was performed at 850°C for 1 hour. At first low temperature GaN interlayer was grown on nitridated sapphire at 575°C at a rate of 6.3 nm/min. Then at 730 °C GaN layers were grown at 7.4 nm/min. Throughout the GaN growth the reflection high-energy electron diffraction (RHEED) pattern remained streaky, indicating the 2D growth mode (not shown here).

Under the same growth conditions, two samples were grown varying the thickness of GaN epitaxial layers, namely B17 and B18 with thicknesses ~120 nm and ~1μm respectively as shown in Fig. 2. Both samples have been annealed at four different temperatures such as

650°C, 725°C, 800°C and 875°C using Rapid thermal processing (RTP) system (AnnealSys-AS-ONE 150) in nitrogen ($N_2$) ambient for 5 mins. Now, as grown and the annealed samples were studied in terms of structural, electrical, optical and surface morphological properties in order to realize the quality of the films.

XRD measurements were performed around symmetric (0002) peak of GaN thin films. Rocking curve ω scan for that particular peak was performed to study FWHM and screw dislocation density. Surface morphology was studied by optical microscope (Olympus) and AFM. PL measurements are performed using a He-Cd laser of excitation power (19 mW) at 325 nm as the excitation source. Electrical properties of the as grown and annealed samples were studied by Hall Effect measurement system.

3. Results and Discussions:

X-ray rocking curve measurements were carried out to study the structural properties of the as grown GaN as well as the annealed GaN epilayers. FWHM of the ω scan around symmetric (0002) peak of the as grown B17 sample was found to be 0.287°. But, when it is annealed at 725°C, the same had reduced to 0.225°. Further increase in annealing temperature gave rise to enhancement in FWHM. Therefore, it can be concluded that sample B17 annealed at 725°C exhibits better structural quality. Similar trend also has been seen in case of the sample B18. Screw type dislocation density from the ω scan around (0002) peak has been calculated using the following relation [24]

$$\rho_{screw} = \frac{\Delta\omega_{(0002)}^2}{9b_s^2} \quad (1)$$

Where $\rho_{screw}$ is the screw dislocation density and $b_s$ is the length of the burger vector ($b_s$ = 0.5185 nm [25]). Screw dislocation density for the as grown B17 sample and the B17 samples annealed at temperatures 650°C, 725°C, 800°C and 875°C were measured to be 1.04 × $10^9$ cm$^{-2}$, 1.17 × $10^9$ cm$^{-2}$, 6.38 × $10^8$ cm$^{-2}$, 7.2 × $10^8$ cm$^{-2}$ and 9.27 × $10^8$ cm$^{-2}$ respectively. For B18 the same are calculated to be 7.69× $10^8$ cm$^{-2}$, 7.99 × $10^8$ cm$^{-2}$, 6.86 × $10^8$ cm$^{-2}$, 6.75 × $10^8$ cm$^{-2}$, 7.15 × $10^8$ cm$^{-2}$. For both B17 and B18, lowest screw dislocation densities were obtained for samples annealed at 725°C (6.4 × $10^8$ cm$^{-2}$ and 6.9 × $10^8$ cm$^{-2}$ respectively). Variation of (a) FWHM of ω scan around GaN (0002) peak and (b) screw type dislocation density with different annealing temperatures for the two samples are shown in Fig. 2.

Hall measurements were performed using LakeShore Model 8404 AC/DC Hall Effect Measurement System (HMS) at room temperature in order to compare the electrical properties of the GaN thin films. Prior to the experiment, the Point contacts were made at each corner of the samples and the size of the contacts was approximated to be near about 1/10th of the distance between two adjacent contacts. Then Van der Pauw factors ($f_{VP}$) for the as grown samples B17 and B18 have been found to be 0.97 and 0.75 respectively. Electron mobilities and carrier concentrations measured for the as grown B17 using the same $f_{VP}$ factor mentioned above were 33cm$^2$/V.sec and 3.3×$10^{18}$ cm$^{-3}$. And the same for the as grown B18 were 39 cm$^2$/V.sec and 1.15×$10^{18}$ cm$^{-3}$ respectively. This clearly explains that with increase in thickness from B17 to B18, reduction in carrier concentrations leading to a relatively higher mobility in B18 has been noticed. Mobility and carrier concentrations for the sample B17 at the annealing temperature 725°C have been found to be 21.4 cm$^2$/V.sec and 9.58 ×10$^{18}$ cm$^{-3}$ and those for the sample B18 the values were 75 cm$^2$/V.sec and 0.83×10$^{18}$ cm$^{-3}$. When annealed at 725°C, B18 has showed improvement in mobility and carrier concentrations but mobility in case of B17 reduces due to increase in carrier concentrations. The variation of annealing temperature on mobility and carrier concentrations is pictured in Fig. 3.

Several studies [26,27] have been already made on the growth temperature dependent surface morphology of the GaN films. We report here the influence of annealing temperatures on surface morphology of all GaN epitaxial thin films performed by Asylum Research AFM system using SPM controller. RMS roughness of the as grown B17 sample is found to be 1.5 nm and the pits density calculated by the use of WsXM softwate is 1.6 × 10$^9$ cm$^{-2}$. But when the sample is annealed to 725°C, roughness as well as pits density are found to be 1.7 nm and 5.1 × 10$^8$ cm$^{-2}$ respectively. So, no improvement in the rms roughness even after annealing at 725°C has been found but one order reduction in pits density has been observed. For the sample B18, as grown GaN thin films have rms roughness and pits density 3.3 nm and 2.0×10$^9$ cm$^{-2}$ respectively, whereas the sample annealed at 725°C shows roughness and the pits density to be 1.8 nm and 1.6×10$^9$ cm$^{-2}$ respectively. In this case quality improvement has been noticed in terms of both rms roughness and pits density between the as grown and the sample annealed at 725°C. Therefore, better surface morphology is achieved for the GaN films annealed at 725°C. Moreover, it can be concluded from the as grown samples that with the increase in the GaN film thickness, rms roughness and pits density increase. Surface images of the as grown and the annealed GaN films at 725°C for both the samples obtained by AFM have been shown in figure 4(a) to 4(d) and rms roughness and pits density calculated using WsXm software have been shown in 4(e) and 4(f) respectively.

Room temperature (RT) PL measurement was performed in order to analyse the optical properties of the as grown as well as the annealed samples. FWHM of the near band edge (NBE) emission (which is attributed to the neutral donor bound excitons) peak of as grown GaN thin films grown on sapphire substrates have been compared with that of the annealed samples. FWHM for the as grown B17 has been found to be 35 meV, while had reduced to 30 meV after annealing it at 725°C. But for B18, FWHM of as grown sample has been found to be 58 meV, whereas that for the annealed one at 725°C was 50 meV. Change in (a) FWHM and (b) NBE peak position with as grown and four different annealing temperatures for both samples have been shown in Fig. 5. For unstressed bulk GaN films, the value of the NBE peak was taken to be 3.40 eV [28]. Hence, the blue shift has been calculated for B17 as grown and the samples annealed at four different temperatures to be 10.9, 10.8, 9, 13.1, 9.9 meV respectively with respect to strain free GaN films, whereas for B18 red shift has been observed for as grown and the sample annealed at 800°C (1.6 and 8.6 meV respectively) in addition to blue shift for other annealing temperatures. Blue shift for B18 annealed at 725°C has been found to be 0.9 meV. The shift observed for all the samples is expected to be due to the presence of strain effects.

4. Summary:

The GaN films, annealed at 725°C ex-situ, grown on intermediated GaN layer/sapphire substrates presents lower value of the FWHM of (0002) XRD peak and hence lower value of screw dislocation density ($6.4\times10^8$ cm$^{-2}$). For surface morphology, AFM provides smaller RMS value and good surface flatness. RT Hall measurements carried out on GaN films indicates increase in electron mobility upon annealing (38.7 cm$^2$/V.sec (as grown) to 75 cm$^2$/V.sec (annealed at 725°C)). RT PL measurements show minimum FWHM value of 30 meV for the annealing temperature of 725°C. The above results indicate that 725°C may be considered as the optimum annealing temperature to have better quality GaN epitaxial films on sapphire substrate.

Acknowledgement:
The work is supported by the Centre of Excellence in Nanoelectronics-phase II (CEN-phase II) and one of the authors, Mahitosh Biswas, would like to thank Indian Nanoelectronics User Program (INUP) sponsored by DeitY and MCIT, Govt. of India for providing the financial support for carrying out the research work.

**Figure Captions:**

1. Schematic diagram of (a) B17 and (b) B18.

2. (a) FWHM values of ω scan around (0002) peak using double axis diffactometer for B17 & B18 (for as grown and annealed samples). Inset shows the ω scan for both the as grown samples.(b) Screw dislocation density ($\rho_{screw}$) as a function of annealing temperatures for samples B17 & B18.

3. Variation of mobility and carrier concentration with annealing temperatures for the sample (a) B17 and (b) B18.

4. Surface morphology taken by AFM over 3×3 μm$^2$ areas for (a) B17 as grown, (b) B17 annealed at 725°C, (c) B18 as grown, and (d) B18 annealed at 725°C. Variation of rms surface roughness and pits density as a function of annealing temperature for both (e) B17 and (f) B18.

5. (a) Variation of PL FWHM of NBE peak of GaN with annealing temperatures for the samples B17 and B18. (b) NBE GaN peak shift as a function of annealing temperatures for both the samples.

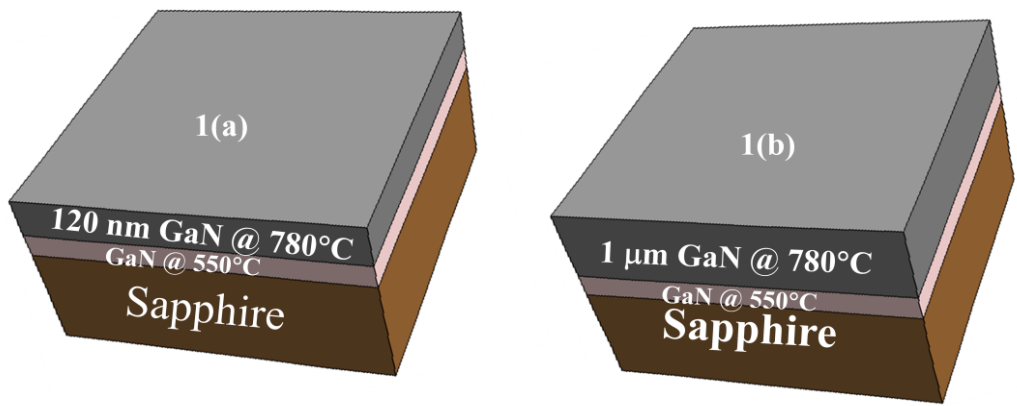

Fig. 1

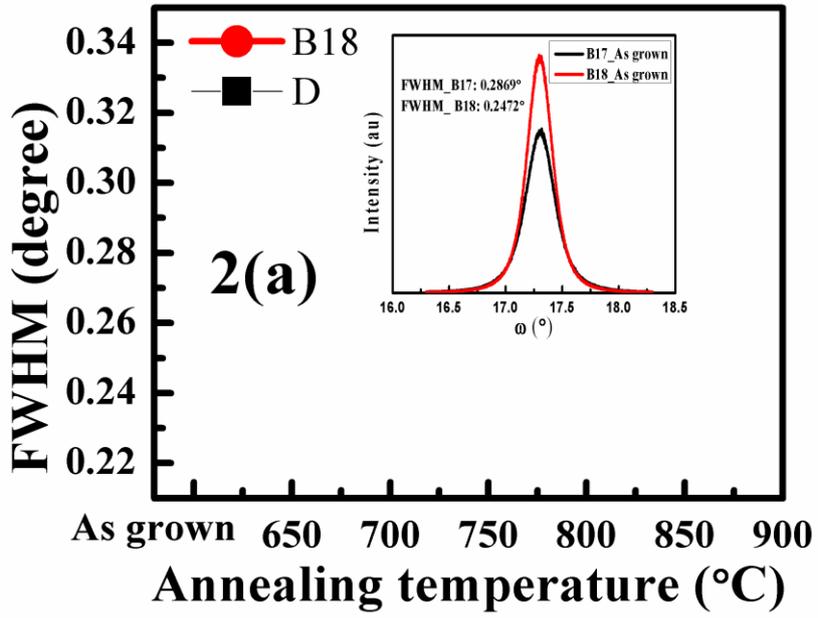

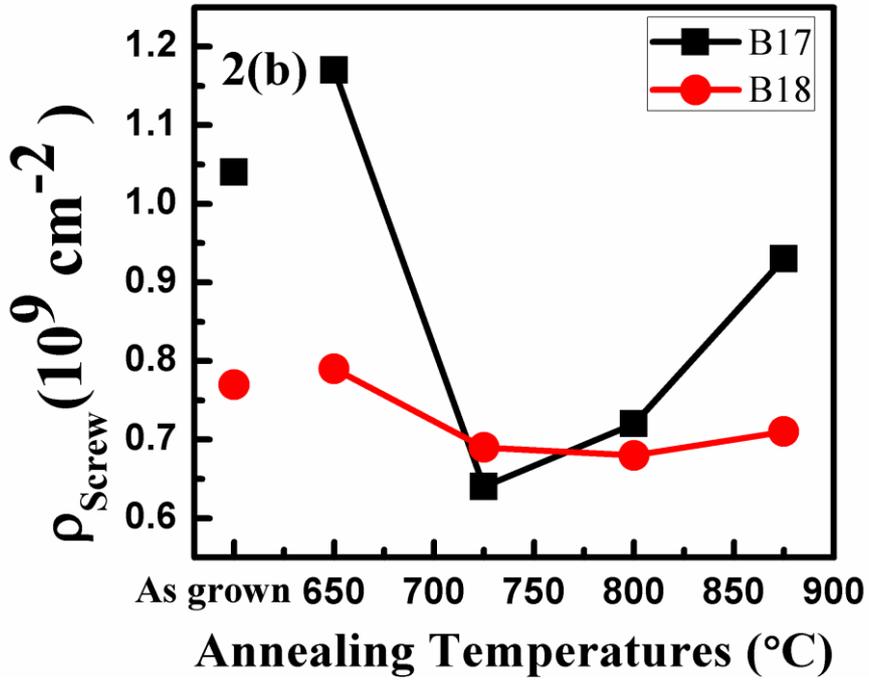

**Fig. 2**

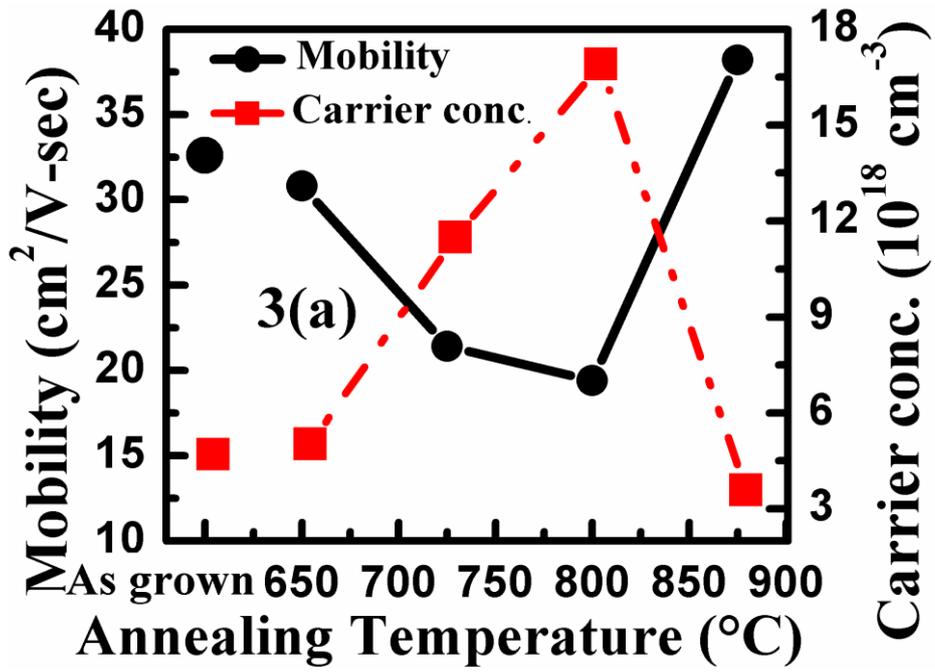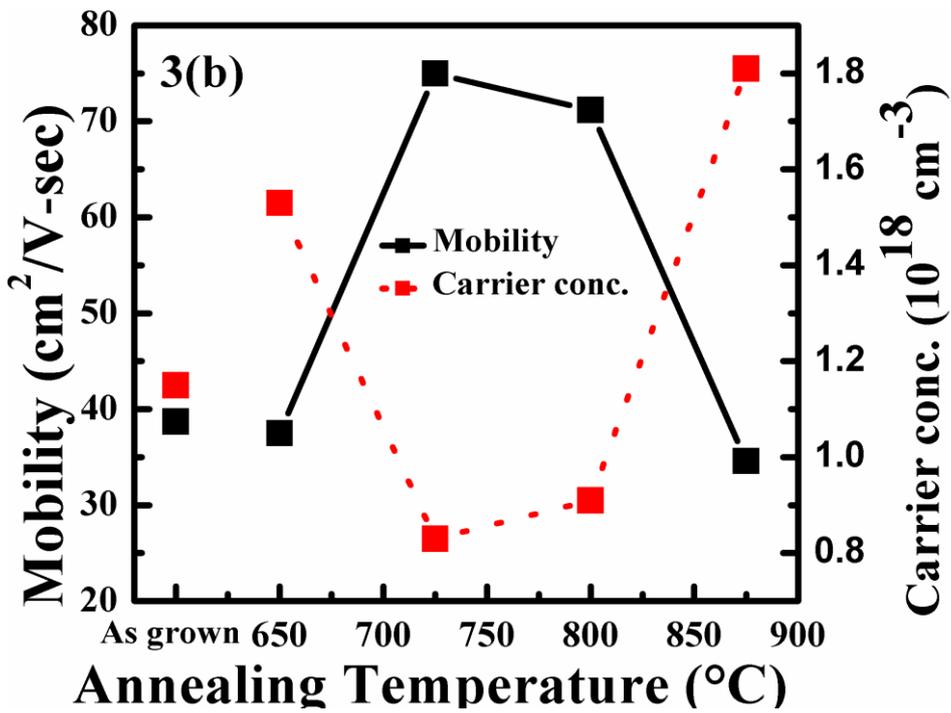

Fig. 3

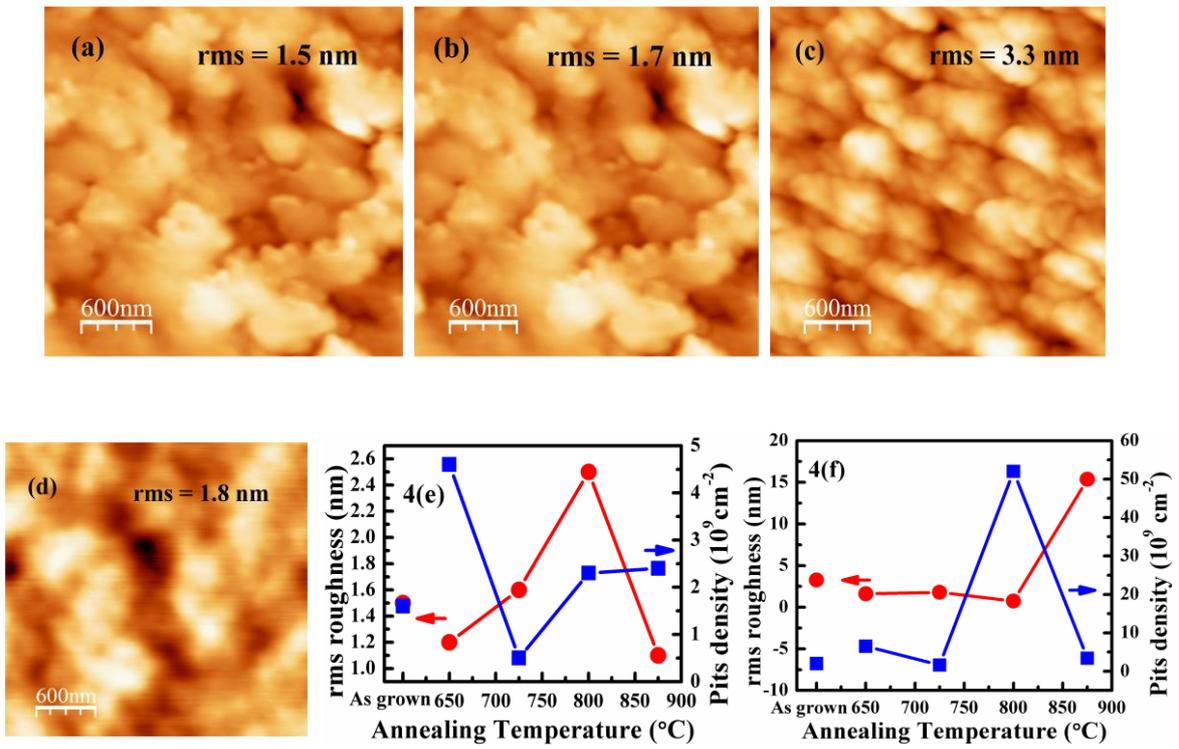

Fig. 4

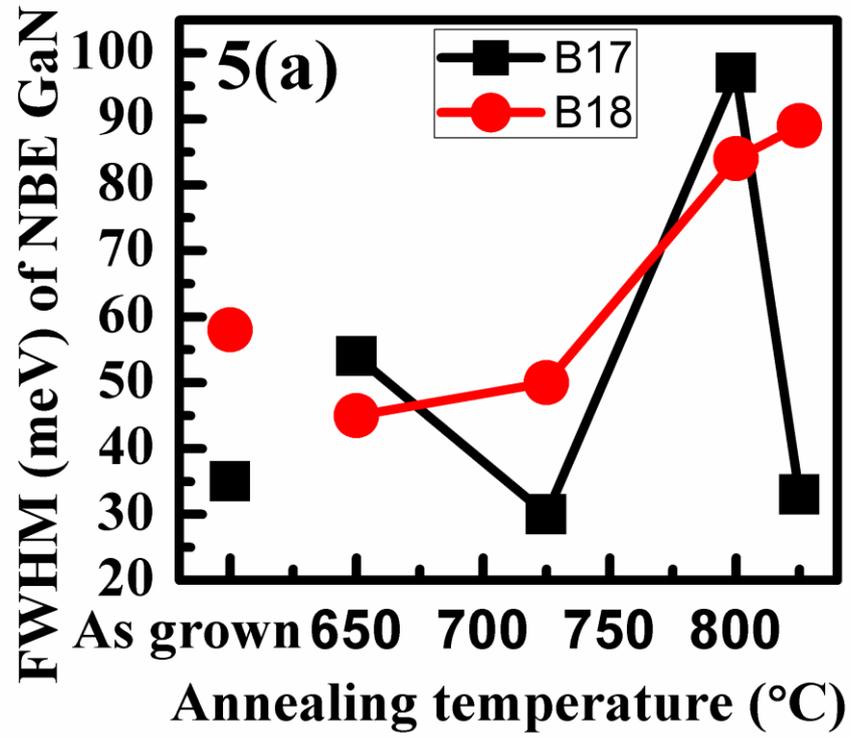

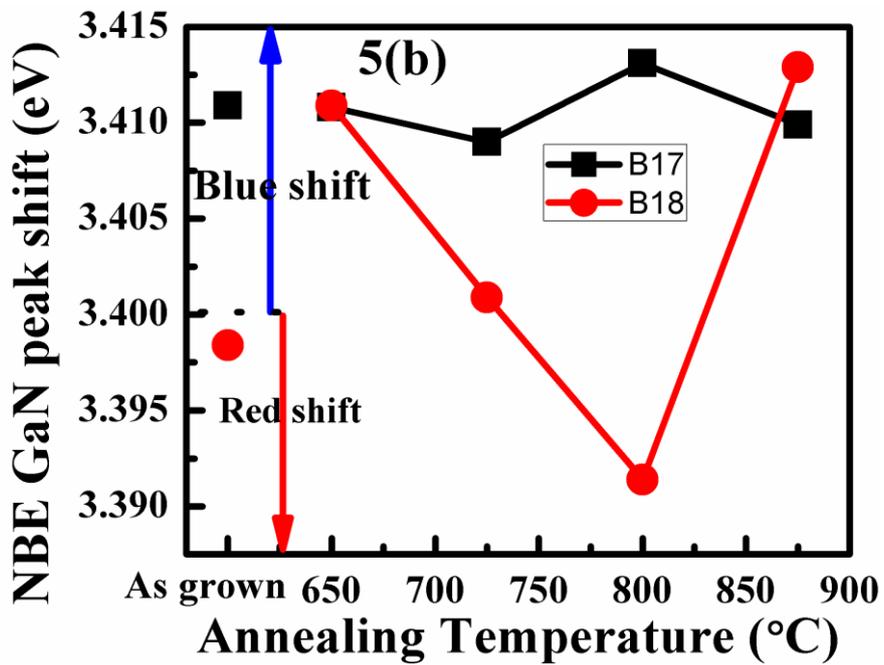

Fig. 5